# Period-doubling bifurcation of dissipative-soliton-resonance pulses in a passively mode-locked fiber laser

YUFEI WANG,[1,2] LEI SU,[2] SHUAI WANG,[3] LIMIN HUA,[3] LEI LI,[3] DEYUAN SHEN,[3] DINGYUAN TANG,[4] ANDREY KOMAROV,[5] MARIUSZ KLIMCZAK,[6,7] SONGNIAN FU[1], MING TANG[1], XIAHUI TANG[1], AND LUMING ZHAO,[1,2,3,*]

[1]*School of Optical and Electronic Information, Huazhong University of Science and Technology*

[2]*School of Engineering and Materials Science, Queen Mary University of London, London E1 4NS, UK*

[3]*Jiangsu Key Laboratory of Advanced Laser Materials and Devices, Jiangsu Collaborative Innovation Center of Advanced Laser Technology and Emerging Industry, School of Physics and Electronic Engineering, Jiangsu Normal University, Xuzhou, 221116 Jiangsu, China*

[4]*School of Electrical and Electronic Engineering, Nanyang Technological University, Singapore*

[5]*Institute of Automation and Electrometry, Russian Academy of Sciences, Academician Koptyug Prospekt 1, 630090 Novosibirsk, Russia*

[6]*Faculty of Physics, University of Warsaw, Warsaw, Poland*

[7]*Institute of Electronic Materials Technology, Wólczynska 133, 01-919 Warsaw, Poland*

*\* corresponding author: lmzhao@ieee.org*

**Abstract:** We report on the experimental observation of period-doubling bifurcation of dissipative-soliton-resonance (DSR) pulses in a fiber laser passively mode-locked by using the nonlinear optical loop mirror. Increasing the pump power of the fiber laser, we show that temporally a stable, uniform DSR pulse train could be transformed into a period-doubling state, exhibiting two sets of pulse parameters between the adjacent cavity roundtrip. It is found that DSR pulses in the period-doubling state could maintain the typical feature of DSR pulse – fixed pulse peak power and linear variation in pulse width with respect to the pump power change. The mechanism for achieving period-doubling of DSR pulses is discussed.

## 1. Introduction

Passively mode-locked fiber lasers have been widely used in various fields today because of their easily manageable structure and unparalleled performance. The study of pulse dynamics in fiber lasers is always a hot topic in recent decades. On the way of pursuing larger pulse energy, dissipative-soliton-resonance (DSR), a novel existence of pulse formation was discovered by Chang. et al in 2008 [1]. Based on numerical simulations of the cubic-quintic complex Ginzburg-Landau equation (CGLE), they found that there existed a parameter regime where the mode-locked pulse could maintain its peak power and linearly broaden its pulse width with the increase in the gain. Owing to the clamped pulse peak power, the pulse energy of the DSR pulse could be boosted arbitrarily without pulse breaking. In 2009, Wu. et al first experimentally demonstrated the DSR pulse generation in a fiber laser by utilizing the nonlinear polarization rotation technique in the normal dispersion regime [2]. Following that, researchers further investigated the DSR pulse as a universal phenomenon that could be obtained under different cavity structures such as the figure-of-eight [3], figure-of-nine [4] and linear cavity [5], with various kinds of mode-locking techniques such as using the nonlinear optical loop mirror (NOLM) [6], using the nonlinear amplifying loop mirror [7] and using novel materials [8–10]. Recently, DSR-based fiber lasers have shown their great potential as a reliable source to generate pulses with ultra-high pulse energy. Semaan et al reported a single DSR pulse with 10 μJ pulse energy directly from a figure-of-eight double-clad Er:Yb co-doped fiber laser [11]. Combining with the master oscillator power amplifier (MOPA), Zheng et al reported a DSR pulse with 0.33 mJ pulse energy by using a figure-of-nine double-clad Tm-doped fiber laser in the 2 μm band [12].

As a promising pulse energy boost mechanism, it is necessary to investigate the dynamics and characteristics of DSR pulses in order to further optimize the performance of DSR-based fiber lasers. It is expected that DSR pulses should have many common features of solitons in the nonlinear system. There have been many reports on characteristics of DSR pulses. For example, multi-pulse states [13–15], harmonic mode-locking [16,17] and the compressibility of DSR pulses [18]. We noticed that DSR pulses among all the above-mentioned researches temporally exhibit a uniform pulse train. As it is well-known, nonuniformity of a soliton pulse train caused by period bifurcation could be an intrinsic feature of all mode-locked fiber lasers [19]. Period bifurcation, which is also known as periodic pulsations, usually can be observed in the time domain where the output of the pulse train would show periodic changes in pulse parameters with the cavity roundtrips. Experimental observation of period-doubling and -tripling bifurcation of mode-locked pulses in a fiber laser was firstly reported by Tamura et al. [20]. Since then, period bifurcation has experimentally been observed in fiber lasers with different scenarios of soliton pulses, such as dispersion-managed solitons [21], multiple solitons [22], bound solitons [23], gain-guided solitons [24] and vector solitons [25]. Theoretically, Akhmediev et al found the parameter regime of the CGLE where various periodic pulsating solitons could exist [26]. Thus, these results brought us a hypothesis that whether the DSR pulses could also exhibit the characteristics of period bifurcation, showing a nonuniform pulse train. The clamped peak power of DSR pulses seems to be naturally resistant to the formation of period bifurcation state exhibiting the change in the peak power [27]. To the best of our knowledge, no periodic behaviors of DSR pulses have yet been reported.

In this paper, we reported the experimental observation of period-doubling bifurcation of DSR pulses in an Yb-doped all-normal-dispersion fiber laser. For the first time, we demonstrated the process of a uniform DSR pulse train transforming into a period-doubling bifurcation state with the increase in the pump power. Different from the conventional behavior of the increase in the pulse peak difference between two adjacent pulses [20], DSR pulses in a period-doubling state maintain their pulse peak power while their pulse width linearly increases with respect to the pump power increase. To achieve the period-doubling of DSR pulses, the pulse peak power of DSR pulses should be able to be increased in practice.

## 2. Experimental setup

The experimental setup is a figure-of-eight fiber laser as shown in Fig. 1. The fiber laser is mode locked by using a NOLM. The cavity is constructed by a 30/70 optical coupler (OC) connecting a unidirectional ring and a loop mirror. The intensity of propagating light in the loop mirror is modulated by the 30/70 OC. Therefore, the transmission of the loop mirror will act as a sinusoidal function, initializing the mode-locking [6]. The 30-cm-long ytterbium-doped fiber (YDF) is pumped by a 600 mW, 976 nm laser diode (LD) through a wavelength division multiplexer (WDM). 20% of the intra-cavity power is output by a 20/80 OC. A polarization independent isolator is used to ensure the counterclockwise propagation of light in the unidirectional ring. To carefully adjust the polarization state of the laser cavity, we employ the three-paddle PCs in both unidirectional ring and the loop mirror. To increase the nonlinearity, we insert a 200-m-long single mode fiber into the loop mirror. All fiber components in this fiber laser work in the normal dispersion regime.

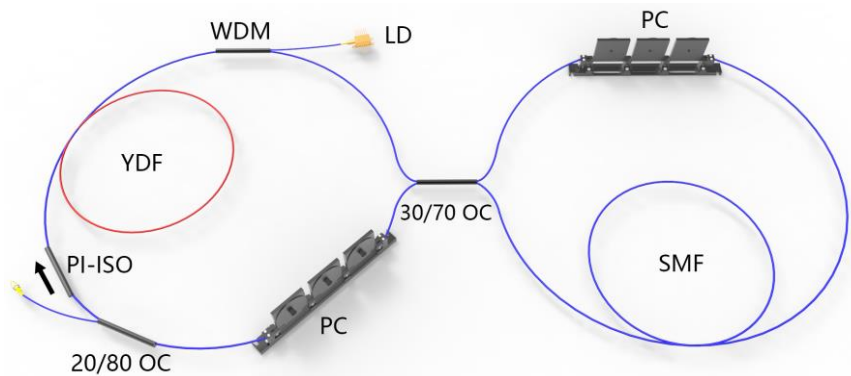

Fig.1 Experimental setup of the fiber laser.

## 3. Experimental results and discussion

Firstly, we observed the generation of DSR pulses in the laser cavity. When the pump power reached the mode-locking threshold, mode-locked pulses at nanosecond-scale could rise from the continuous wave by appropriately setting the two PCs. Figure 2(a) shows the temporal trace of pulses monitored by using a high-speed photodetector (New focus 1014, 45-GHz) in combination with a digital oscilloscope (Agilent DSA-X 96204Q, 63-GHz). The response time of the monitor system is ~20 ps. Started with pump power of 55.7 mW, the generated pulse showed a square-shaped temporal profile, with a pulse width of 631 ps. Further increasing the pump power, the trailing edge of the pulse broadened while the peak of pulse remained almost constant due to the peak power clamping (PPC) effect [28]. Figure 2(b) plots the change in temporal pulse width and average output power of the fiber laser when the pump power was increased from 55.7 to 129.5 mW. The pulse width increased linearly from 631 ps to 1.597 ns, and the average output power gradually increased from 0.18 to 0.496 mW. The linear change in these two parameters indicates that the calculated peak power of the pulse remained almost constant during the pulse broadening. The results showed the fingerprint of DSR pulses. Figure 2(c) shows the radio frequency (RF) spectrum detected by an analyzer (Agilent N9320B) with 1 MHz span and 10 Hz resolution bandwidth (RBW) when the pump power was increased to 129.5 mW. The first order of the RF spectrum peaked at 900 kHz, which corresponded to the cavity fundamental repetition frequency. The signal-to-noise ratio increased from 60.6 dB when the pump power was 55.7 mW to 69.5 dB when the pump power was 129.5 mW. With the 10 MHz span and 1 kHz RBW, the trace of RF spectrum showed a uniform degradation as plotted in Fig. 2(d). The RF spectrum confirms that the single DSR pulse operated in a stable state.

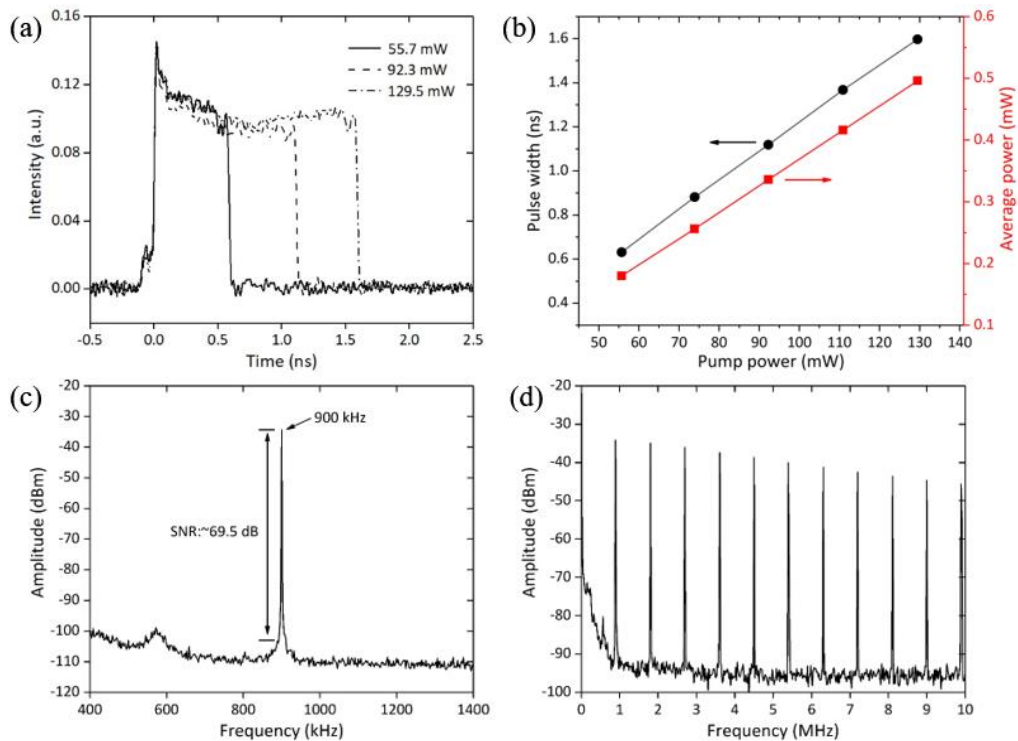

Fig. 2 (a) DSR pulse profiles and (b) pulse width and average output power change versus pump power; (c) RF spectrum with span of 1 MHz; (d) RF spectrum with span of 10 MHz.

Figure 3(a) shows the oscilloscope trace of temporal pulse train with the pump power of 129.5 mW. A uniform pulse train can be observed as pulses with equal intensity repeatedly showed up every cavity roundtrip. The period is ~1.11 μs, in agreement with the reciprocal of the cavity fundamental repetition rate. With fixed settings of PCs, the uniform state of the pulse train could be maintained when we increased the pump power only. However, when the pump power was increased to 260.6 mW, the intensity of the pulse started to have a periodic difference. As shown in Fig. 3(b), although the pulse was still mode-locked at the cavity period, the pulse intensity varies vividly every two cavity roundtrips. It suggested that the DSR pulse exhibited the property of period-doubling bifurcation in the time domain.

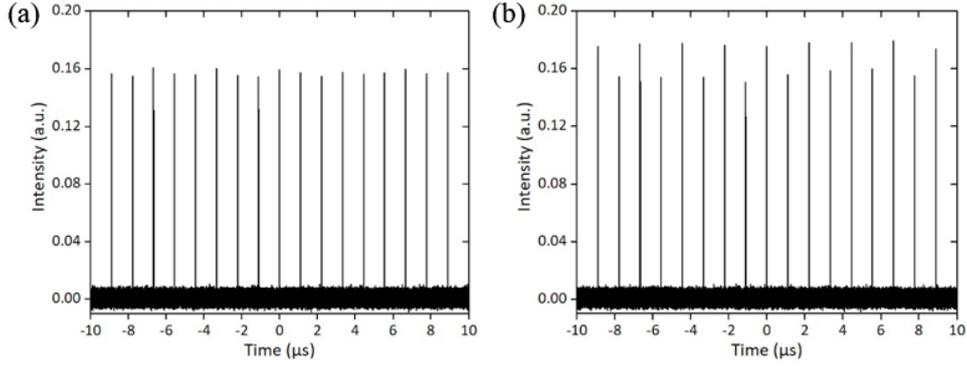

Fig. 3 The oscilloscope trace of the DSR pulse train at (a) period-one state; (b) period-two state

When the period-doubling bifurcation occurs, new frequency component will simultaneously appear on the RF spectrum [21]. To confirm the generation of period-doubling bifurcation, we measured the RF spectrum of the pulse train with the pump power of 260.6 mW. Figure 4(a) plots the RF spectrum with 10 MHz span and 1kHz RBW. Compared with Fig. 2(d), there were newly generated frequency components locating at the half of the cavity fundamental repetition frequency and its higher order harmonics. Figure 4(b) plots the detailed RF spectrum with 1 MHz span and 10 Hz RBW. The first order of the RF spectrum jumped from 900 kHz to 450 kHz. The pulse train we obtained at the pump power of 260.6 mW was operating in a so-called period-two state.

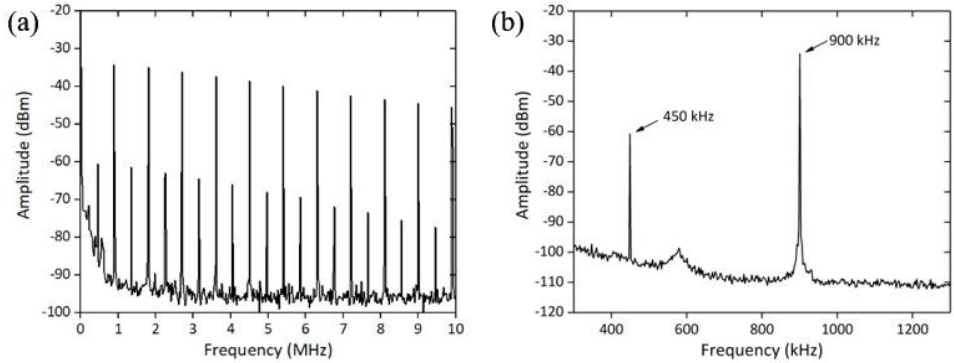

Fig. 4 The RF spectrum of period-two state with (a) 10 MHz span; (b) 1 MHz span.

It is well-known that the appearance of period-doubling is one of the intrinsic nonlinear properties of fiber lasers and it is a threshold effect [29]. The nonlinearity in a fiber laser could be identified by the accumulated nonlinear phase shift during the pulse propagation in a fiber laser, which is proportional to the nonlinear coefficient, the pulse peak power, and the propagation length [30]. For a fiber laser, the pulse peak power generally increases with respect to the increase in the pump power, therefore enhancing the accumulated nonlinear phase shift. Consequently, the appearance of period-doubling is possible. For a fiber laser generating DSR pulses, as the pulse peak will be kept constant during the pump power increase, it is out of our expectation for the appearance of period-doubling.

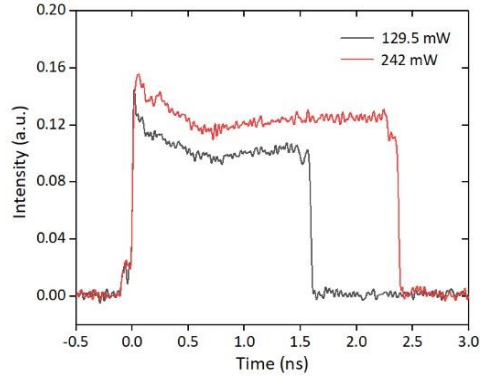

Fig. 5 Pulse profiles of the DSR pulse at pump power of 129.5 mW and 242 mW, respectively.

To get insight of the generation of period-doubling of DSR pulses, detailed experimental analyses were carried out. We were back to Fig. 2 and found that there indeed exists a slight pulse peak power increasing with the pump power. The pulse peak power increased from 0.317 W to 0.345 W when the pump power increased from 55.7 mW to 129.5 mW. Therefore, even under the DSR operation, it is still possible to increase pulse peak power while the pulse peak power is supposed to be clamped. However, this kind of pulse peak power improvement is slight and limited. It is almost impossible to achieve the stepping over the threshold requirement for the generation of period-doubling. Recently we found that in a DSR pulse fiber laser, it is possible to achieve DSR pulse breaking, DSR pulse harmonic mode locking, or DSR pulse narrowing by solely increasing the pump power [31]. It is resulted from the multiple parameter changes caused by solely increasing the pump power. Theoretically, the solely increase in the gain will result in the fingerprint behavior of DSR pulses – linearly broadening of pulse width with fixed pulse peak power [1,18]. However, in practice, the sole increase in the pump power may cause multiple parameter changes apart from the expected gain increase only [31], leading to the change in pulse profile. Experimentally we measured the pulse profile when we increased the pump power from 129.5 mW to 242 mW. As shown in Fig. 5, it is found that both the width and the peak of the pulse increased while the pulse train is still under period-one state. The obvious change in pulse peak indicates that the DSR pulse no longer maintained its linear broadening characteristics within this pump power regime. Further increasing the pump power, as long as the pulse peak power can be improved to fulfill the nonlinearity threshold requirement, period-doubling of DSR pulses can be achieved in the fiber laser as shown in Fig. 3(b).

It is interesting to know the dynamics of DSR pulses operated in period-doubling state. After the pulse train was transformed into the period-doubling state, we recorded the detailed temporal traces of the adjacent high- and low-intensity pulses. We found that the PPC effect may continue to play a role after the period-doubling operation was achieved. In another word, the DSR performance could be maintained even after the period-doubling operation is achieved. With another fixed settings of PCs, we found a pump power regime where the pulses exhibiting the period-doubling state showed typical characteristics of DSR pulse broadening. As shown in Fig. 6(a) and Fig. 6(b), the pulse profiles of the high-/low- intensity pulse suggest that DSR pulses experiencing period-doubling had same pulse width at fixed-level of pump power. The pulse width was temporally broadened from 6.827 to 8.134 ns when the pump power was increased from 360.1 to 394.5 mW. Linear pulse parameter changes and PPC effect can be retrieved from Fig. 6(c) and Fig. 6(d), which clearly demonstrates the DSR performance. We note that the DSR broadening process under the period-doubling state only exists within a very narrow pump power tuning range of ~35 mW here. Larger pump power will make the DSR pulse unstable.

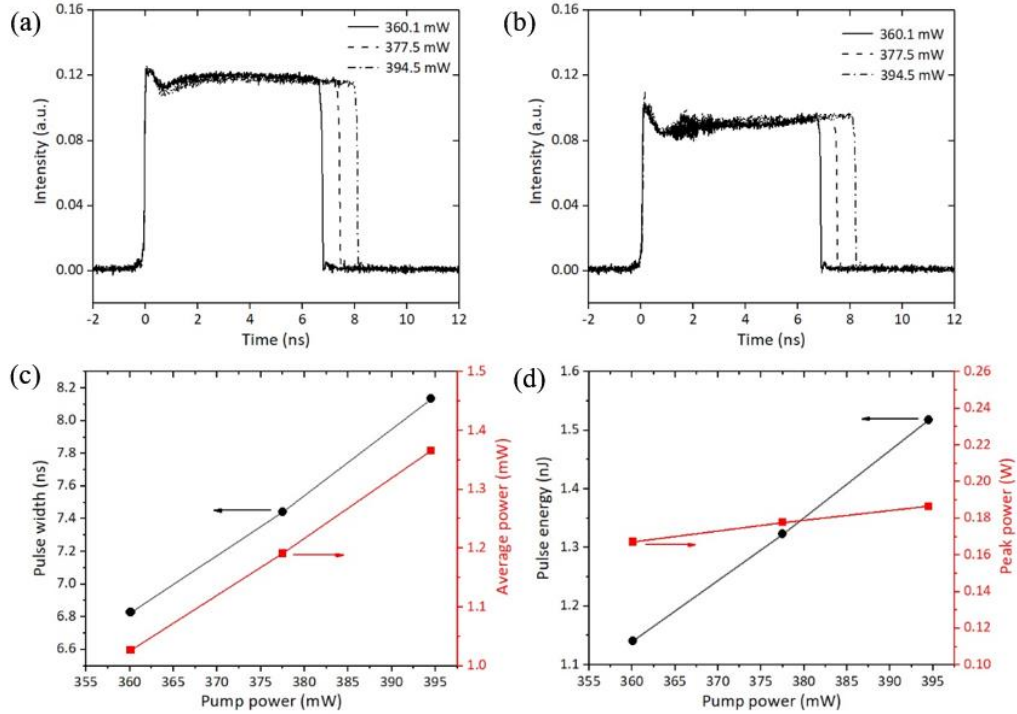

Fig.6 DSR pulse broadening in the period-doubling state with increasing pump power. (a) High-intensity pulse profile; (b) Low-intensity pulse profile; (c) Pulse width and average output power versus pump power; (d) Pulse energy and pulse peak power versus pump power.

Experimentally we also observed the phenomenon of DSR pulse narrowing under period-doubling state. As shown in Fig. 7, the pulse width of both the high-intensity pulses and the low-intensity one reduced when the pump power was increased from 260.6 to 297.4 mW. The pulse widths of both pulses started to shrink while the peak of the pulse remained constant as the pump power increased. The pulse narrowing phenomenon of a DSR pulse has been reported previously, which is caused by the gain competition in the laser [31]. We note that the high and low intensity pulse always had the same pulse width during the pulse narrowing. The evolution suggests that the pulses exhibiting period-doubling behavior can show the characteristics of DSR pulses no matter it is a DSR broadening or narrowing. The pulse would become unstable when the pump power was beyond 297.4 mW.

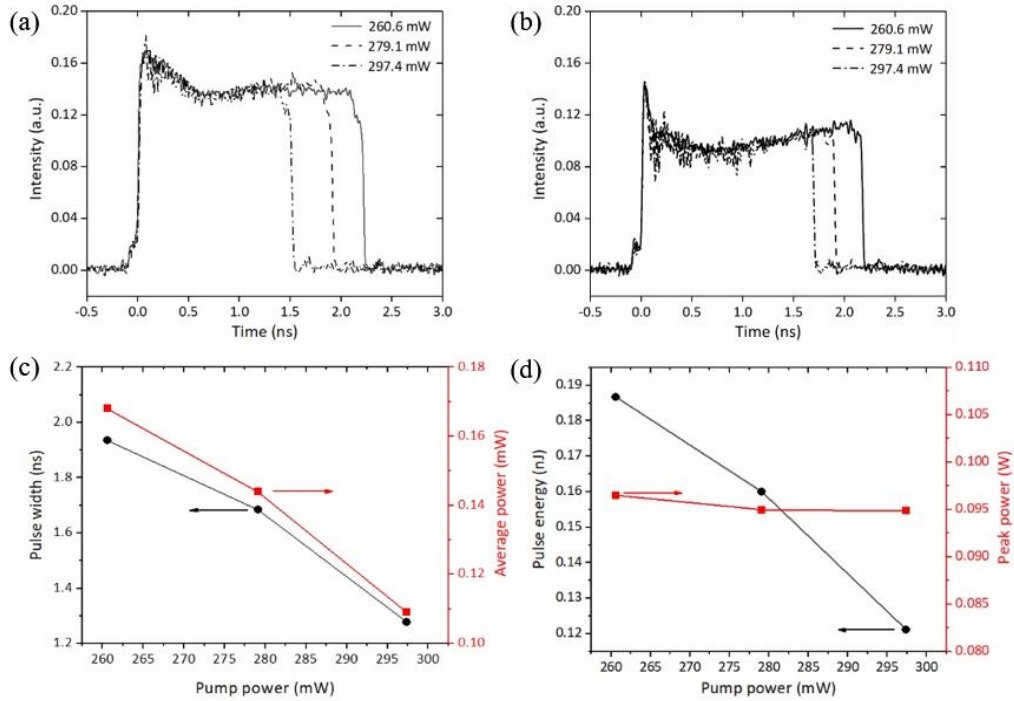

Fig. 7 DSR pulse narrowing in the period-doubling state with increasing pump power. (a) High-intensity pulse profile; (b) Low-intensity pulse profile; (c) Pulse width and average output power versus pump power; (d) Pulse energy and pulse peak power versus pump power.

## 4. Conclusion

In conclusion, we have experimentally observed the period-doubling bifurcation of DSR pulses in a fiber laser. DSR pulses can transform into the period-doubling state by simply increasing the pump power. Temporally, DSR pulses in the period-doubling state had two sets of the pulse parameters in the adjacent cavity roundtrip. Our experimental observations confirm that in a fiber laser, the pulse can simultaneously exhibit the phenomena of DSR and period-doubling bifurcation. It was shown that DSR pulses exhibiting period-doubling could maintain their peak during the process of pulse broadening or pulse narrowing with respect to the pump power increase. The findings enrich the dynamics of DSR pulses, which will benefit the design of ultrafast lasers with desirable soliton patterns for practical applications.

## Funding

National Natural Science Foundation of China (NSFC) (11674133, 11911530083, 61575089); Russian Foundation for Basic Research (RFBR) (19-52-53002); Royal Society (IE161214); Protocol of the 37th Session of China-Poland Scientific and Technological Cooperation Committee (37-17); European Union's Horizon 2020 research and innovation programme under the Marie Skłodowska-Curie grant agreement No. 790666. We acknowledge support from Jiangsu Overseas Visiting Scholar Program for University Prominent Young & Middle-aged Teachers and Presidents and Priority Academic Program Development of Jiangsu Higher Education Institutions (PAPD). Mariusz Klimczak acknowledges support from Fundacja na rzecz Nauki Polskiej (FNP) in scope of First TEAM/2016-1/1 (POIR.04.04.00-00-1D64/16).